\documentstyle[12pt,float]{article}

\title{Study of Strain and Temperature Dependence of Metal Epitaxy}

\author{C. Ratsch, P. Ruggerone, and M. Scheffler\\
{\it Fritz-Haber-Institut der Max-Planck-Gesellschaft, Faradayweg 4-6,}\\
{\it D-14195 Berlin-Dahlem, Germany}}

\begin{document}

\maketitle

\section{INTRODUCTION}\label{intro}

Metallic films are important in catalysis, magneto-optic storage media, and
interconnects in microelectronics, and it is crucial to predict and control
their morphologies. The evolution of a growing crystal is determined by the
behavior of each individual atom, but technologically relevant structures have
to be described on a time scale of the order of (at least) tenths of
a second and on a length scale of nanometers. 
\begin{table}[b]
\vspace{.5cm}
\caption{
The time and length scales handled by different theoretical approaches to 
study growth.
}
\begin{tabular}{lccc}
\hline 
& Type of information & Time scale & Length scale \\
\hline 
Density functional  & Microscopic & - & $\leq 10^3$ atoms\\
theory calculations & & &\\ \hline
Ab initio  & Microscopic &  $t \leq $ 10 ps & $\leq 10^2$ atoms \\ 
molecular dynamics & & & \\ \hline
Semi-empirical & Microscopic & $t \leq $ 1 ns & $\leq 10^3$ atoms \\ 
molecular dynamics & & & \\ \hline
Kinetic Monte Carlo & Microscopic & 1 ps $\leq t \leq$ 1 hour 
& $\leq$ 1 $\mu$m \\
& to Mesoscopic & & \\ \hline
Rate equations & Averaged & 0.1 s $\leq t \leq \infty$ & All \\ \hline
Continuum equations & Macroscopic & 1 s $\leq$ t $\leq \infty$ & $\geq$ 10 nm
\\ 
\hline
\end{tabular}
\label{approaches}
\end{table} 
An adequate theory of growth should describe the atomistic level
on very short time scales (femtoseconds), the formation of small islands
(microseconds), as well as the evolution of mesoscopic and macroscopic
structures (tenths of seconds).  Different theoretical approaches to the
description of growth phenomena address different time and length scales, and
an overview is given in Table~\ref{approaches}.  These different techniques
should be regarded as complementary to each other rather than as alternatives,
and we will show in this article how some of them may successfully be
combined.

The development of efficient algorithms combined with the availability of
cheaper and faster computers has turned density functional theory (DFT) into a
reliable and feasible tool to study the microscopic aspects of growth
phenomena (and many other complex processes in materials science, condensed
matter physics, and chemistry).  These calculations provide the so-called
potential energy surface (PES, also often called total energy or
Born-Oppenheimer surface) that is the potential energy experienced by a
particle during chemical reactions and diffusion. The basic concepts behind
DFT are sketched in Section~\ref{sec:concepts}. A single microscopic process
on such a PES can be characterized by either doing a molecular dynamics (MD)
simulation or by using transition state theory.  Both approaches typically
will give the same result. Problems may occur when the energy barrier is
comparable to the thermal energy and correlated effects cannot be neglected.
An overview over the different processes and how they can be described within
transition state theory is given in Section~\ref{sec:atomistic}.

Technologically relevant applications of growth often involve disparate
materials with different lattice constants. However, it is not understood so
far how strain in a system influences the most fundamental process that occurs
during epitaxial growth, the diffusion of a single adatom on a substrate. It
is the aim of this article to discuss the current understanding of this
aspect. In Section~\ref{sec:strain} some recent results for the strain
dependence of surface diffusion that is determined by an energy barrier and a
prefactor will be presented.  While the DFT results were obtained for only a
few metallic systems, we assert that the physical trends discussed are rather
general.

A DFT total energy calculation by its nature is primarily a {\it static}
calculation.  An accurate way to describe the {\it dynamical} evolution of a
growing crystal is given by MD simulations. It is not possible to carry out
such a simulation on mesoscopic and/or macroscopic time and length scales.  Ab
initio MD runs usually can cover at most times of picoseconds, whereas
semi-empirical MD runs may extend up to some nanoseconds (at the cost of the
accuracy). This is because an MD time step cannot be longer than the inverse
of a typical phonon frequency. For example, a MD run simulates the entire
sequence of unsuccessful attempts occurring between two successful diffusion
events that may be separated by interval of the order of nanoseconds.  Thus,
MD can model only very few events, and a proper statistics cannot easily be
obtained for growth processes. Moreover, since growth patterns usually develop
on a time scale of seconds, the inadequacy of MD is evident. Finally, the
growth structures involve large numbers of particles ($\sim 10^{2}$ to
$10^{4}$), far beyond the reach of MD (simulations with $\sim 10^2$ atoms are
hardly feasible, and only for very short times).  Because an adequate
treatment of the statistical nature of the problem is typically not achieved,
we believe that the value of MD is often overstated, and the method of choice
for studying the spatial and temporal development of growth is kinetic Monte
Carlo (KMC).

The growing crystal defines a lattice system onto which the non-crystalline
growth units (atoms or molecules) are deposited. In the case of growth from
solution or from vapor one may assume with sufficient generality that the
growth units impinge on the surface (or, more exactly, on the solid-vapor)
interface at random with an average frequency. The same random character
belongs to the diffusion of growth units on the surface. It is therefore
natural to describe deposition, desorption, and diffusion
mechanisms as stochastic processes, and this matches very well with the
stochastic nature of the Monte Carlo method.  A brief introduction to this
method and its combination with microscopic parameters obtained from {\it ab
  initio} parameters is given in Section~\ref{sec:kmc}.  It is shown that
realistic {\it ab initio} kinetic Monte Carlo simulations are able to predict
an evolving mesoscopic structure on the basis of microscopic details.

In order to compare quantitatively results from a Monte Carlo simulations with
experimentally measured quantities, such as for example the density of islands
on the surface, one needs to take a statistical average over this quantity. In
a KMC simulation this is obtained by doing a simulation on a sufficiently
large lattice.  An alternative and more elegant approach is the construction
of mean field equations that by definition only contain averaged quantities.
An example are rate equations discussed briefly in Section~\ref{sec:concepts}.
Numerical solution of these equations can give the density of islands, but
without microscopic details.  This ansatz currently is not sufficiently
advanced to make reliable predictions for experimentally relevant quantities
and needs to be developed further.  As the final goal a proper mean field
approach should yield the same results as the statistical average over the
results of a more detailed microscopic approach.  The inclusion of more
details (that in principle is possible) unfortunately makes a mean field or
continuum approach less tractable, so that new techniques have to be
developed.

\section{BASIC CONCEPTS OF DENSITY\\ FUNCTIONAL THEORY}
\label{sec:concepts}

The total energy of an $N$-electron, poly-atomic system is given by the
expectation value of the many-particle Hamiltonian using the many-body
wave-function of the electronic ground state.  For a solid or a surface the
calculation of such an expectation value is impossible within the framework of
a wave-function approach. However, Hohenberg and Kohn~\cite{hoh64} have shown
that the ground-state total energy can also be obtained without explicit
knowledge of the many-electron wave-function by minimizing an energy
functional $E[n]$.  This is the essence of density-functional theory (DFT),
which is primarily (though in principle not exclusively) a theory of the
electronic ground state, couched in terms of the electron density $n({\bf r})$
instead of the many-electron wave-function $\Psi(\{{\bf r}_i\})$.

The important theorem of Hohenberg and Kohn~\cite{hoh64} (see also
Levy~\cite{lev79}) states: The specification of a ground state density $n({\bf
  r})$ determines the corresponding external potential $v^{\rm ext}({\bf r})$
$uniquely$ (to within an additive constant),
\begin{equation}
n({\bf r}) \to v^{\rm ext}({\bf r})\quad .
\label{defrel}
\end{equation} 
The external potential $v^{\rm ext}({\bf r})$ is typically (and
definitely for our purpose here) the Coulomb potential due to the nuclei,
assumed at fixed positions as they are much heavier than the electrons
(Born-Oppenheimer approximation).
While the other direction [$v^{\rm ext}({\bf r}) \to n({\bf r})$] is well
known to exist (because $v^{\rm ext}({\bf r})$ determines the many-particle
Hamiltonian $H$) Eq. (\ref{defrel}) is less obvious.  The theorem enables to
transform the variational problem of the many-particle Schr\"odinger equation
into a variational problem of an energy functional:
\begin{equation}
E_0 \leq \langle \Psi|H|\Psi \rangle = E_v[\Psi[n]] = E_v[n]\quad ,
\label{variat}
\end{equation} 
with $E_0$ the energy of the ground state, and $E_v[n] =
\int d{\bf r}\, v^{\rm ext}({\bf r}) n({\bf r}) + G[n]$.  The variable is 
$n({\bf r})$ (the electron density of any
$N$-electron system), and $v^{\rm ext}({\bf r})$ is kept fixed.  $G[n]$ is a
{\em universal} functional independent of the system, i.e., independent of
$v^{\rm ext}({\bf r})$.  
The main advantage of this approach is that
$n({\bf r})$ only depends on three variables, while $\Psi(\{{\bf r}_i\})$
depends on all the coordinates of the $N$ electrons~\cite{spin}.
Thus, it is plausible that the variational problem of $E_v[n]$ is
easier to solve than that of $\langle \Psi|H|\Psi \rangle$, yet the result for
the ground-state energy and the ground state electron density will be the
same. The total energy is~\cite{units}
\begin{equation}
E^{\rm tot} ( \{{\bf R}_J \} ) = E_0( \{ {\bf R}_J \} ) +
\frac{1}{2}\sum_{J,J',J \ne J'} \frac{Z_J Z_{J'}}{|{\bf R}_J -
{\bf R}_{J'}|} \quad,
\label{tot}
\end{equation} 
where $\{ {\bf R}_J\}$ includes all atoms, and $Z_J$ is the
nuclear charge.

An important problem remains, namely that the exact form of the functional
$G[n]$ is unknown. Earlier work (in particular the Thomas-Fermi approach) had
shown that the treatment of the kinetic energy $\langle \Psi |
-\frac{1}{2}\nabla^2 | \Psi \rangle$ is of particular importance and Kohn and
Sham~\cite{koh65} therefore wrote the energy functional in the form
\begin{equation} 
E_v[n]  =  T_s[n] + \int d{\bf r}\, v^{\rm ext}({\bf r}) n({\bf r}) + 
\frac{1}{2}  \int d{\bf r}\,  v^{\rm H}({\bf r}) n({\bf r}) + E^{\rm xc}[n] \quad ,
\label{excdef}
\end{equation} 
where $T_s[n]$ is the functional of the kinetic energy of a
system of non-interacting electrons with density $n({\bf r})$, and $v^{\rm
  H}({\bf r}) = \int d{\bf r'}\, \frac{n({\bf r'})}{|{\bf r} - {\bf r'}|}$ is
the Hartree potential that describes the electrostatic interaction between
electrons.  $E^{\rm xc}[n]$, the so-called exchange-correlation functional, 
accounts for the Pauli principle, dynamical correlations due to the Coulomb
repulsion, and the correction of the self-interaction included for convenience
in the Hartree term. With Eq. (\ref{excdef}) the problem of the unknown
functional $G[n]$ is mapped onto one that involves $T_s[n]$ and $E^{\rm
  xc}[n]$. Although the functional $T_s[n]$
is not known explicitly in a mathematically closed form, it can be evaluated
exactly by using the following ``detour'' proposed by Kohn and Sham. The
variational principle applied to Eq. (\ref{excdef}) leads to
\begin{equation}
\frac{\delta E_v[n]}{\delta n({\bf r})} = 
\frac{\delta T_s[n]}{\delta n({\bf r})} + v^{\rm eff}({\bf r}) = \mu \quad ,
\label{variational2}
\end{equation}
where $\mu$ is the Lagrange multiplier associated with the requirement 
of a constant particle number and thus equals the electron chemical
potential. The effective potential is defined as 
\begin{equation}
v^{\rm eff}({\bf r}) = v^{\rm ext}({\bf r}) + v^{\rm H}({\bf r}) +
v^{\rm xc}({\bf r}) \quad,
\label{eff_pot}
\end{equation} 
with $v^{\rm xc}({\bf r}) = \delta E^{\rm xc}[n] / \delta
n({\bf r})$, and $n({\bf r})$ is a ground-state density of any non-interacting
electron system, i.e.,
\begin{equation}
n({\bf r}) = 2\sum_{i=1} f_i | \phi_i({\bf r}) |^2 \quad,
\label{denans}
\end{equation} 
where we introduced the occupation numbers $f_i$. The factor
2 accounts for the spin degeneracy. Since $T_s[n]$ is the kinetic energy
functional of non-interacting electrons, Eq. (\ref{variational2}) [together
with Eq. (\ref{denans})] is solved by
\begin{equation}
\left[- \frac{1}{2} \nabla^2 + v^{\rm eff}({\bf r}) \right] \phi_i({\bf r}) = 
\epsilon_i \phi_i({\bf r}) \quad.
\label{kseq}
\end{equation} 
These are the Kohn-Sham equations, that are to be solved
self-consistently together with Eqs. (\ref{eff_pot}) and (\ref{denans}).  In
principle, this gives the exact ground-state electron density and total energy
of a system of interacting electrons.  However, the functional $E^{\rm xc}[n]$
is still unknown. Some general properties of this functional and values for
some special cases are known.  Detailed and very accurate understanding exists
for systems of constant electron density. 
These results for $\epsilon^{\rm xc}:= \epsilon^{\rm
xc}_{\rm LDA}(n)$ are then used in the functional
\begin{equation}
E^{\rm xc}_{\rm LDA} [n] = \int d{\bf r} \, n({\bf r}) 
\epsilon^{\rm xc}_{\rm LDA} (n({\bf r}))\quad ,
\label{lda.exc}
\end{equation}
which is the local-density approximation (LDA)~\cite{koh65}.
Thus, in the LDA the many-body effects are included such that for
a homogeneous electron gas the treatment is exact and for
an inhomogeneous system exchange and correlation are treated by
assuming that the system can be composed from many small systems with a
locally constant density.

The LDA can be improved by including the dependence on the density gradient
which leads to the generalized gradient approximation (GGA).  Several
different GGA's were proposed in the
literature~\cite{per92,per86,heval,bec88,lee88} and have been used
successfully for DFT calculations for atoms, molecules, bulk solids, liquids,
and surfaces (an overview can be found in Refs.~\cite{bec96,per96}), but also
limitations have been pointed out~\cite{mit94,umr96}. It is by now clear that
the lattice constants calculated with a GGA are typically larger than those
obtained with the LDA, with the experimental values usually being in between.
Binding energies (or cohesive energies) of molecules and solids are clearly
improved by the GGA as well as energy barriers of chemical reactions.  (see
Ref.~\cite{ham94} and references therein). A general rule of thumb for the use
of LDA or GGA is still missing and tests should be carried out for each
specific problems. The total energies are changed when going from the LDA to
the GGA, thus for surface diffusion the changes in energy barriers, i.e., in
total energy {\em differences}, may be pronounced as well although the rank of
the processes is not altered (see Refs.~\cite{yu96,rat97}). The GGA data are
usually more reliable, but still careful calculations with various forms of
GGA must be performed.

The general method described above can be transformed into an operative tool 
following diverse schemes that essentially differ in the basis employed to expand
the electronic wave function. A widely used basis set (also chosen for
the calculations presented below) consists of plane waves 
(see Refs.~\cite{Payne,bock97}) but other choices are possible~\cite{other1,other2}. 
A deeper discussion of the implementations of the DFT plane-wave basis set method for
the study of growth phenomena can be found in 
Ref.~\cite{woodru97}.

\section{ATOMISTIC PROCESSES}
\label{sec:atomistic}

\begin{figure}[tb]
\unitlength1cm
 \begin{center}
    \begin{picture}(10,6.5)
      \includegraphics{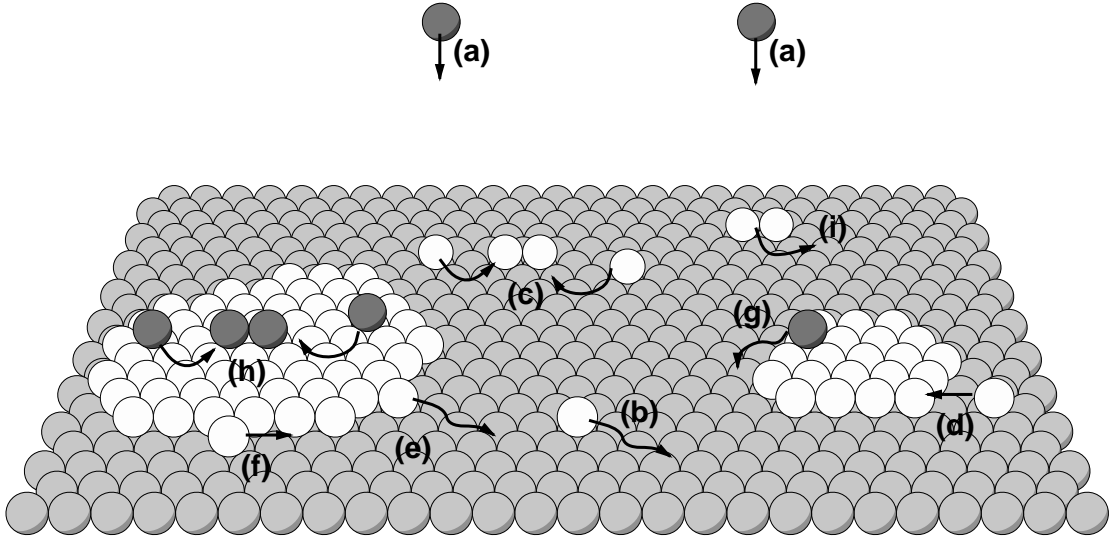}
    \end{picture}
 \end{center}
\caption{
The different atomistic processes for adatoms on a surface. See the text for a 
discussion.}
\label{processes}
\end{figure}

The different atomistic processes encountered by adatoms that 
are deposited onto a surface are illustrated in
Fig.~\ref{processes}.  After deposition $(a)$ atoms can diffuse across the
surface $(b)$ and will eventually meet another adatom to form a nucleus
$(c)$ or get captured by an already existing island or a step edge $(d)$. Once
an adatom has been captured by an island, it may either break away from the
island ({\it reversible aggregation}) $(e)$ or remain bonded to the island
({\it irreversible aggregation}). An atom that is bonded to an island may
diffuse along its edge $(f)$ until it finds a favorable site.  As long as the
coverage of adsorbed material is low (say $\Theta \le 10~\%$), deposition on
top of islands is insignificant and nucleation of islands on top of existing
islands practically does not occur. However, if the step down motion $(g)$ is
hindered by an additional energy barrier, nucleation of island on top of
islands $(h)$ becomes likely at some critical coverage.
In this article we will mainly focus on the migration of adatoms
on the flat terrace or along step edges and will not discuss effects that are
due to mass transport between different layers.  Moreover, we will say nothing
about the possibility of long jumps even though it most likely plays a role
for certain systems as it has been observed with the field ion microscope for
Pd on W\,(211) by Senft and Ehrlich~\cite{sen95}, and very recently in a scanning
tunneling microscopy study for Pt on Pt\,(011) by Linderoth {\it et
  al.}~\cite{lin97}.

\subsection{Transition State Theory}

To understand the diffusion of an adatom on a surface 
we need to calculate its potential-energy surface (PES): 
\begin{equation}
E^{\rm PES}(X_{\rm ad}, Y_{\rm ad}) = \min_{Z_{\rm ad},\{{\bf R}_I\}}
E^{\rm tot} (X_{\rm ad}, Y_{\rm ad}, Z_{\rm ad},\{{\bf R}_I\}) \quad,
\label{PES}
\end{equation} 
where $E^{\rm tot} (X_{\rm ad}, Y_{\rm ad}, Z_{\rm ad},\{{\bf
  R}_I\})$ is the ground-state energy of the many-electron system (also
referred to as the total energy) at the atomic configuration $(X_{\rm ad},
Y_{\rm ad}, Z_{\rm ad},$ $\{{\bf R}_I\})$.  According to Eq. (\ref{PES}), the
PES is the minimum of the total energy with respect to the $z$-coordinate of
the adatom $Z_{\rm ad}$ and all coordinates of the substrate atoms $\{{\bf
  R}_I\}$.  Assuming that vibrational effects can be neglected, the minima of
the PES represent stable and metastable sites of the adatom.  Note that this
PES refers to slow motion of nuclei and assumes that for any atomic
configuration the electrons are in their respective ground state. Thus, it is
assumed that the dynamics of the electrons and of the nuclei are decoupled
(Born-Oppenheimer approximation).

The dynamics of an adatom on such a PES can 
be described by different approaches.
In a molecular dynamics simulation the forces on each atom 
are computed and the atoms are moved accordingly.
Such a simulation is very computer intensive, and we will discuss the 
advantages and disadvantages in more detail in Section~\ref{sec:kmc}.
An alternative idea that will be used in this article is that 
processes such as diffusion are
described by rates. 
Within transition state theory (TST)
the rate of a microscopic process $j$ usually has the form~\cite{gla41,vin57,wah90}
\begin{equation}
\Gamma^{(j)}= \frac{k_{\rm B}T}{h} \exp(-\Delta F^{(j)}/k_{\rm B}T)\quad ,
\label{ratdef}
\end{equation}
where $\Delta F^{(j)}$ is the difference in the Helmholtz free energy
between the maximum (saddle point) and the minimum
(equilibrium site) of the potential curve along the reaction path of
the process $j$. $T$ is the temperature, $k_{\rm B}$ the Boltzmann
constant, and $h$ the Planck constant. The free energy of activation 
$\Delta F^{(j)}$ needed by the system to move from the initial position to 
the saddle point is given by
\begin{equation}
\Delta F^{(j)} = E_{\rm d}^{(j)} - T \Delta S_{\rm vib}^{(j)}\quad .
\label{deltaf}
\end{equation}
Here $E_{\rm d}^{(j)}$ is the sum of the differences in the static total and
vibrational energy of
the system with the particle at the minimum and at the saddle point,
and $\Delta S_{\rm vib}^{(j)}$ is the analogous difference in the
vibrational entropy. The rate of the process $j$ can be cast as
follows:
\begin{equation}
\Gamma^{(j)} = \Gamma_0^{(j)} \exp(-E_{\rm d}^{(j)}/k_{\rm B}T)\quad ,
\label{retdef1}
\end{equation} 
where $\Gamma_0^{(j)} = (k_{\rm B}T/{h}) \exp (\Delta S_{\rm
  vib}^{(j)}/k_{\rm B})$ is the effective attempt frequency. In the case of
isotropic motion of an adatom on the surface it follows from
Eq. (\ref{retdef1}) that the diffusion constant is $D = D_0 \exp(-E_{\rm
  d}^{(j)}/k_{\rm B}T)$~\cite{bob}. The prefactor $D_0 = 1/(2\alpha)
\Gamma_0^{(j)} l^2$, where $l$ is the jump length and $\alpha$ the
dimensionality of the motion ($\alpha =2$ for the surface).

The two basic quantities in Eq. (\ref{retdef1}) are the attempt frequency
$\Gamma_0^{(j)}$ and the activation energy $E_{\rm d}^{(j)}$.
TST~\cite{vin57,wah90} allows an evaluation of $\Gamma_0^{(j)}$ within the
harmonic approximation:
\begin{equation}
\Gamma_0^{(j)} = \frac{\prod_{i = 1}^{3N} \, \nu_i}{\prod_{i = 1}^{3N-1} \,
  \nu_i^{*}}\quad ,
\label{harmappr}
\end{equation} 
where $\nu_i$ and $\nu_i^{*}$ are the normal mode frequencies
of the system with the adatom at the equilibrium site and at the saddle point,
respectively, and $3N$ is the number of degrees of freedom. The denominator in
Eq. (\ref{harmappr}) contains the product of only $3N-1$ normal frequencies,
because for the adatom at the saddle point one of the modes describes the
motion of the particle toward the final site and has an imaginary frequency.
TST is only valid when $E_{\rm d}^{(j)}$ is larger than $k_{\rm B}T$.

The attempt frequency $\Gamma_0^{(j)}$ shows a much weaker temperature
dependence than the exponential and for typical growth temperatures it is of
the order $10^{12} - 10^{13}s^{-1}$.  When the barriers for two diffusion
events are different a {\it compensation effect} may occur, i.e.,
$\Gamma_0^{(j)}$ is larger for processes with a higher energy barrier.
Indeed, a higher energy barrier usually implies a larger curvature of the
potential well around the equilibrium site of the adatom. The corresponding
vibrational frequencies of the adatom in such a potential are larger as well,
which implies [see Eq. (\ref{harmappr})] that the attempt frequency increases.
Boisvert {\it et al.}~\cite{boi95} found that for metal diffusion the 
prefactor might decrease by up to one order of magnitude for very low 
energy barriers.

\begin{figure}[tb]
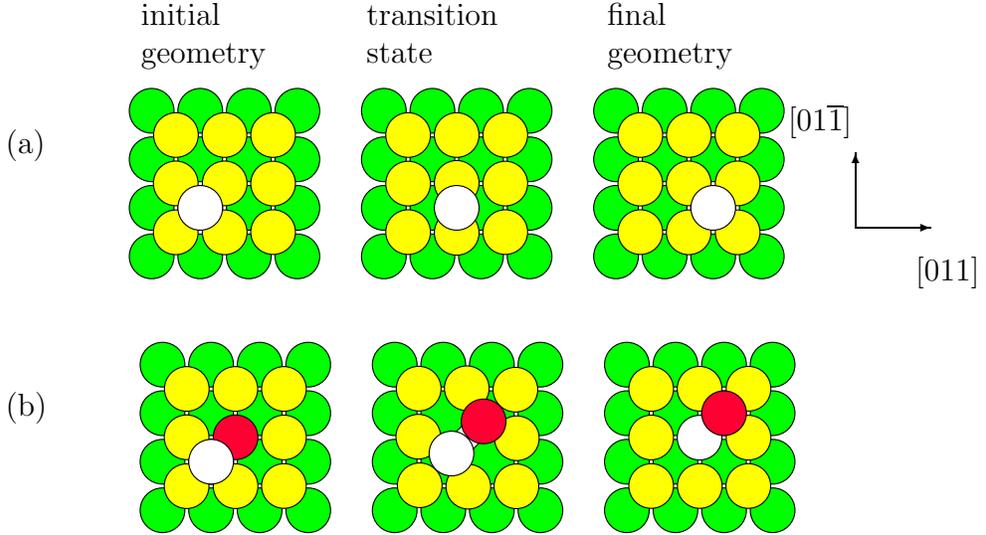

\unitlength1cm
\begin{center}
   \begin{picture}(10,10.0)
      \includegraphics{hop100f}
      \includegraphics{exc100f}
\put(-1.0,4.5){(b)}
\put(-1.0,8){(a)}
\put(10.3,7.0){\vector(0,1){1}}
\put(10.3,7.0){\vector(1,0){1}}
\put(9.4,8.3){$[01\overline{1}]$}
\put(11.1,6.3){$[011]$}
\put(0.8,9.7){{initial}}
\put(0.8,9.2){{geometry}}
\put(3.8,9.7){{transition}}
\put(3.8,9.2){{state}}
\put(7.0,9.7){{final}}
\put(7.0,9.2){{geometry}}
   \end{picture}
\end{center}
\vspace{-3cm}
\caption{
  Diffusion via hopping (a) and exchange (b) on a fcc\,(100) surface.}
\label{more_exchange}
\end{figure}

Now consider all possible paths $l$ that describe process $j$ to get from one
stable or metastable adsorption site, ${\bf R}_{\rm ad}^{(j)}$, to an adjacent
one, ${\bf R}_{\rm ad}^{(j)}{\bf '}$.  The energy difference $E_{{\rm
    d}l}^{(j)}$ between the energy at the saddle point along $l$ and the
energy of the initial geometry is the barrier for this particular path. If the
vibrational energy is negligible compared to $E_{{\rm d}l}^{(j)}$, the
diffusion barrier $E_{\rm d}^{(j)}$ then is the minimum value of all $E_{{\rm
    d}l}^{(j)}$ of the possible paths that connect ${\bf R}_{\rm ad}^{(j)}$
and ${\bf R}_{\rm ad}^{(j)}{\bf '}$, and the lowest energy saddle point is
called the {\it transition state}. Although often only the path with the most
favorable energy barrier is important, it may happen that several paths exist
with comparable barriers or that the PES consists of more than one sheet (e.g.
Ref.~\cite{kley96}).  Then the {\it effective} barrier measured in an
experiment or a molecular dynamics (MD) simulation represents a proper average
over all possible pathways.

The common view for surface diffusion is that an adatom hops from one lattice
site to a neighboring lattice site. This mechanism is called {\it hopping}
and for the (100) surface it is illustrated in
Fig.~\ref{more_exchange}(a).  Diffusion might also occur with a very
different mechanism, the so-called {\em diffusion by atomic exchange} (or {\it
  exchange mechanism}).  The adatom can replace a surface atom and the
replaced atom then assumes an adsorption site. This was first discussed by
Bassett and Webber~\cite{bas78} and Wrigley and Ehrlich~\cite{wri78}. Even for
the crystal bulk exchange diffusion has been discussed~\cite{pand86}. This
mechanism is actuated by the desire of the system to keep the number of cut
bonds low along the diffusion pathway.  On fcc\,(100) surfaces diffusion by
atomic exchange was observed experimentally for Pt and Ir~\cite{exp90}, and
theoretically predicted for Al~\cite{fei90}. The geometry for exchange
diffusion at a fcc\,(100) surface is shown in Fig.~\ref{more_exchange}(b).

If one looks at a diffusion event in an experiment one can often not
distinguish the different mechanisms that were operative, so that an
experimentally deduced quantity is only an averaged quantity over the
different mechanisms. Within TST, one can define an individual rate for each
process, for example for surface diffusion via hopping and exchange. In a
temperature range where several of these mechanisms are important, one has to
take the correct average of them to compare the theoretical values with the
experimental ones~\cite{kley96}.

\subsection{Analysis of Experimental Results}

Before discussing results that were obtained for growth parameters from 
ab initio calculations and comparing them with data obtained from 
experimental measurements, we would like to briefly outline how 
experimental data often are analyzed. For simplicity, we will only 
focus on the surface diffusion constant. 

Processes $(a), (b), (c), (d),$ and $(e)$ of Fig.~\ref{processes} can be
described by phenomenological rate equations of the form
\begin{equation}
{dN_1 \over dt} = \Phi - 2\kappa_1 N_1^2 - N_1 \sum_{j>1} \kappa_j N_j
+ 2 \gamma_2 N_2 + \sum_{j>2} \gamma_j N_j
\label{RE1}
\end{equation}
\begin{equation}
{dN_j \over dt} = N_1 (\kappa_{j-1} N_{j-1} - \kappa_j N_j)
- \gamma_j N_j + \gamma_{j+1} N_{j+1} \quad .
\label{RE2}
\end{equation} 
Such a set of equations constitutes the basis of nucleation theory and
describes the time evolution of the adatom density, $N_1$, and the density of
islands of size $j$, $N_j$, for growth on a flat surface in the submonolayer
regime.  Adatoms are deposited onto the substrate at a rate $\Phi = F \,
\cal{N}$ adatom/s where $F$ is the flux in ML/s and $\cal{N}$ the number of
atoms pro ML.  The second and third term in Eq.~(\ref{RE1}) account for
isolated adatoms being ``lost'' because two adatoms can meet at a rate
$\kappa_1$ to form a new nucleus, or adatoms get captured at a rate $\kappa_j$
by an island of size $j$.  The last two terms in Eq.~(\ref{RE1}) describe
further supply sources of adatoms and are gain terms because dimers may
dissociate and adatoms detach from an island of size $j$ at a rate $\gamma_j$.
Equation~(\ref{RE2}) reflects the fact that the number of islands of size $j$
increases because islands of size $j-1$ grow and islands of size $j+1$ shrink.
The number of islands of size $j$ decreases when islands of size $j$ either
shrink or grow.  Note that no evaporation into the gas phase is included in
Eqs.~(\ref{RE1}) and (\ref{RE2}) (that means that the description in terms of
Eqs.~(\ref{RE1}) and (\ref{RE2}) is appropriate only at not too high
temperatures).

With the assumption that agglomerates of $i^* +1$ and more adatoms are stable
against break-up ($\gamma_j = 0$ for $j > i^*$) one derives the scaling
relation~\cite{Scaling1}
\begin{equation}
N^{\rm is} = C_0 \left(D \over F\right)^{-{i^*/(i^* + 2)}}
\label{scaling}
\end{equation} 
where $N^{\rm is}=\sum_{j>i^*}N_j$ is the island density and
$D$ the diffusion coefficient of an adatom on the flat surface.  The number
$i^*$ is called the size of the critical nucleus. The main steps to derive Eq.
(\ref{scaling}) are as follows: A steady state for the adatom concentration is
assumed, i.e., the left-hand side of Eq. (\ref{RE1}) is neglected. This
means essentially that the number of adatoms deposited equals the number of
atoms captured when all islands on the surface are larger than $i^*$. Eq.
(\ref{RE1}) can then be inserted into Eq. (\ref{RE2}), and the latter is
integrated to obtain Eq.  (\ref{scaling}).  The total island density thus
depends on the ratio of the adatom diffusion constant and the deposition flux.
When the diffusion constant increases, a deposited adatom can sample a larger
area before the next adatom is deposited.  Therefore, the density of island
decreases. The same results from a smaller deposition rate, because the time
between two successive deposition events increases, and thus again the area
sampled by a diffusing adatom is enlarged.

It is widely believed that scaling laws that result from nucleation theory can
be applied straightforward, so that measurements of the island density as a
function of the growth temperature can yield the barrier for surface diffusion
and the prefactor. The results presented in Section~\ref{sec:strain} of this
article show that this bears quite some risks and may lead to incorrect
conclusions. Furthermore, it has been shown~\cite{rat94} that relation
(\ref{scaling}) is problematic for $i^*>1$ because not always a well defined
integer $i^*$ exists. As long as the temperature is small enough (or the
deposition rate is large enough), a dimer is essentially stable and $i^*=1$.
When $D/F$ increases, $i^*$ eventually becomes larger than 1, because atoms
can break away from any island. In particular, it is plausible that an atom
that is singly bonded at the edge of a large cluster is more likely to detach
than an atom that is part of a small cluster with all atoms in the cluster
being highly coordinated. Thus, the quantity $i^*$ should be interpreted as an
effective, averaged quantity, and it is clear that it does not have to take on
an integer value. When $D/F$ is small enough $i^*=1$, and we restrict ourselves
to this regime in the present discussion.

In an STM experiment the island density can be measured as a function of the
temperature.  In a log-log plot of $N^{\rm is}$ as a function as $D/F$ one can
obtain $E_{\rm d}$ from the slope of the data, and $D_0$ from the intercept
with the y-axis using Eq. (\ref{scaling}) (as long as $T$ is low enough so
that $i^*=1$).  This is precisely what is often done in experimental studies.
Thus, we would like to stress that a quantity that is often called an {\it
  experimental value} is in fact only the result of an {\it analysis of
  experimental data}, but not a quantity that is measured directly.  An
experimentally deduced value for a diffusion barrier and prefactor can not
distinguish between different mechanisms such as hopping or exchange, and
therefore only represents an effective value that is an average over the
different mechanisms.

It is important to note that the analysis is based on certain assumptions. For
example, it is possible that not only adatoms, but also clusters diffuse on
the surface. In that case, one can still measure the island density, but since
cluster diffusion is not included in Eqs.~(\ref{RE1}) and (\ref{RE2}), the
scaling law described by Eq. (\ref{scaling}) is not applicable anymore, and
numbers extracted with it are meaningless. An analysis by Villain {\it et al.}
\cite{vil92} predicts that the exponent in Eq. (\ref{scaling}) can increase in
a certain range to $0.5$ (for $i^*=1$) when the diffusion of dimers is
allowed.  A recent Monte Carlo simulation by Kuipers and Palmer~\cite{kui96}
showed that the island density $N^{\rm is}$ decreases significantly when
clusters are allowed to diffuse. A decrease of the island density by one order
of magnitude leads to an error for the prefactor of $(i^*+2)/i^*$ orders of
magnitude.

The quantity $C_0$ in Eq.~(\ref{scaling}) is also not known.  It essentially
depends on the capture numbers $\kappa_j$, and it can be estimated from Monte
Carlo simulations to be between 0.1 and 0.5. Using a self consistent approach
for the $\kappa_j$~\cite{bal94} one gets $C_0 \simeq 0.25$. However, an
uncertainty of $\sim$ 5 for $C_0$ translates also into an uncertainty for the
prefactor of two orders of magnitude (for $i^*$ = 1).  We therefore believe
that particularly results obtained for the prefactor from measured island
densities using Eq.~(\ref{scaling}) can be significantly flawed.  An
improvement of scaling laws that are derived under less restrictive
assumptions remains a challenge for the future.

\section{STRAIN DEPENDENCE OF SURFACE\\ DIFFUSION}
\label{sec:strain}

In this Section we will show how the concepts of DFT outlined above can be
applied to study surface diffusion. In particular, we will concentrate on the
effects of strain on energy barriers and attempt frequencies for metallic
systems. We will point out that the analysis of experimental data based on
simple scaling laws is more complicated than usually believed.

\subsection{Ag on Ag\,(111) and Pt\,(111)}

Growth of Ag on Pt\,(111) and Ag on a thin Ag film on Pt\,(111) has been the focus
of recent studies~\cite{bru95,bru97}.  This system has a lattice
mismatch of $4.2\,\%$ and appears to be a well suited system that can provide
important information about the effects of strain during growth. Brune {\it et
  al.}~\cite{bru95} measured the island densities for Ag on Pt\,(111), Ag on one
monolayer (1 ML) of Ag on Pt\,(111), and Ag on Ag\,(111), and employed Eq.~(\ref{scaling})
to extract the diffusion barriers and prefactors. They obtained $E_{\rm
  d}^{\rm{\mbox{\scriptsize{Ag-Pt}}}} = 157$~meV with $\Gamma_0 = 1 \times
10^{13}$~s$^{-1}$ for Ag on Pt\,(111), $E_{\rm
  d}^{\rm{\mbox{\scriptsize{Ag-Ag/Pt}}}} = 60$~meV with $\Gamma_0 = 1 \times
10^{9}$~s$^{-1}$ for Ag on 1~ML Ag on Pt\,(111), and $E_{\rm
  d}^{\rm{\mbox{\scriptsize{Ag-Ag}}}} = 97$~meV with $\Gamma_0 = 2 \times
10^{11}$~s$^{-1}$ for Ag on Ag\,(111).  In particular for the lowest diffusion
barrier the prefactor was found to decrease by 4 orders of magnitude, which 
the authors of Ref.~\cite{bru95} claim is an unusually high compensation effect.

Not many studies have been devoted to investigate the influence of strain on
surface diffusion. The first thorough study we are aware of has been a
molecular dynamics simulation for Si on Si\,(100) that employed a
Stillinger-Weber potential~\cite{rol92}. In this study it was found that the
barrier for diffusion along the fast channel parallel to the dimer rows is
lowered by approximately $10\%$ for $3\%$ compressive as well as for $2\%$
tensile strain, and that atop the dimer rows (in the same direction) it is
increased by about $10\%$. A general trend can not be seen, which might also
be due to the fact that the reconstructions for semiconductor systems make the
behavior more complicated.  For metallic systems there have been a few
theoretical studies in the last few years that employed different methods with
different degrees of sophistication.

\begin{figure}[tb]
\unitlength1cm
\begin{center}
   \begin{picture}(7,6.5)
      \includegraphics{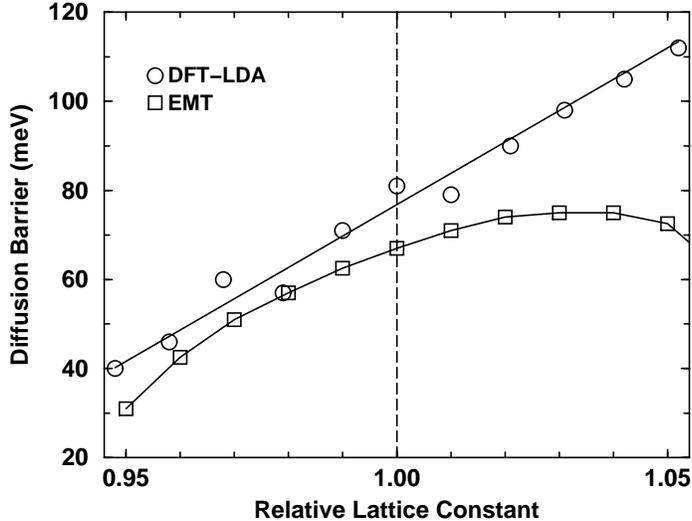}
   \end{picture}
\end{center}
\caption{
  Diffusion barrier (in meV) for Ag on Ag\,(111) as a function of strain.  The
  circles are DFT-LDA results from Ref. \protect{\cite{rat97}} and the squares
  are EMT results from Ref. \protect{\cite{bru95}}.}
\label{Strain_barrier}
\end{figure}

In a recent first principle study~\cite{rat97,rat97_2}
the influence of strain on the diffusion constant for Ag on Ag\,(111) has been 
discussed. It has been found that in the range of $\pm 5\,\%$ strain 
the DFT results for the diffusion barrier exhibit a linear dependence with a
slope of $0.7$ eV as it is displayed in Fig.~\ref{Strain_barrier}.
The calculated diffusion barrier for the unstrained system,
$E_{\rm d}^{\rm{\mbox{\scriptsize{Ag-Ag}}}} = 81$~meV, is in good agreement
(within the error margins of the experiment and the calculations) with the results 
obtained from the analysis of the STM measurements of $E_{\rm
  d}^{\rm{\mbox{\scriptsize{Ag-Ag}}}} = 97$~meV. The accordance between
experiment and theory extends to the system Ag/Pt\,(111) and Ag/1ML
Ag/Pt\,(111). These results are summarized in \mbox{Table
  \ref{Ag_Pt_barriers}.} In Fig.~\ref{Strain_barrier} the DFT-LDA results are
compared to those of an EMT study~\cite{bru95}.  The EMT results exhibit a
linear dependence only for smaller values of strain ($\pm2\,\%$) and the
diffusion barrier starts to decrease for values of misfit larger than $3\,\%$.
\begin{table}[b]
\vspace{.5cm}
\caption{
  Diffusion barriers (in meV) for Ag on Pt\,(111), Ag on one monolayer (ML) Ag
  on Pt\,(111), and Ag on Ag\,(111).}
\begin{tabular}{lcccccc}
\hline
System: Ag on  & \hspace{.0cm} & Experiment \protect{\cite{bru95}} & \hspace{.0cm} &
EMT \protect{\cite{bru95}} & \hspace{.0cm} & DFT \protect{\cite{rat97}} \\
\hline
Pt\,(111) & & 157 &  & 81 & & 150 \\
1ML Ag on Pt\,(111) & & 60 & & 50 & & 65 \\
Ag\,(111) &  & 97 & & 67 & & 81 \\
\hline
\end{tabular}
\label{Ag_Pt_barriers}
\end{table}
Indeed, it is plausible that a decrease of the diffusion barrier occurs when
the atoms are separated far enough that eventually bonds are broken.  However,
as the DFT-LDA results show, for Ag/Ag\,(111) this happens at values for the
misfit that are larger than $5\,\%$.  Additionally, when comparison with
experiment is possible [i.e., Ag on Ag\,(111), and Ag on a monolayer Ag on
Pt\,(111)] the EMT results are off by a factor that varies from 1.2 to 2.

The general trend of an increasing energy barrier for hopping diffusion with
increasing lattice constant is quite plausible (for exchange diffusion see the
next Section).  Smaller lattice constants correspond to a reduced corrugation
of the surface, and as result the atom is not bonded much stronger at the
adsorption sites than at the bridge site.  In contrast, when the surface is
stretched the corrugation increases and the adsorption energy at the
three-fold coordinated hollow sites increases.  In an effective medium theory
study a linear dependence of the surface diffusion constant as a function of
the lattice constant has been found as well for the (111) surfaces of Ni, Pd,
Pt, Cu, Ag, and Au~\cite{mor96}.  The picture will change when the strain is
so large that bonds are broken and then it is expected that the hopping
diffusion barrier will start to decrease again at very large tensile strain.

It is worth noting that the diffusion barrier for Ag on top of a pseudomorphic
layer of Ag on Pt\,(111) is substantially lower than it is for Ag on
Ag\,(111). A question that arises is whether this reduced diffusion barrier is
a result of the compressive strain or should be ascribed to electronic
rearrangements induced by the Pt substrate.  The diffusion barrier for Ag on
Ag\,(111) with a lattice constant that is compressed to the value of the
lattice constant for Pt is $E_{\rm d}^{\rm{\mbox{\scriptsize{Ag-Ag}}}} =
60$~meV while that for Ag on Pt\,(111) (also with the Pt lattice constant of
3.92 \AA \, obtained from DFT) is $E_{\rm
  d}^{\rm{\mbox{\scriptsize{Ag-Ag/Pt}}}} = 65$~meV. The agreement of these two
values suggests that the reduction of the diffusion barrier for Ag on a layer
of Ag on Pt\,(111) is mainly a strain effect and that the diffusion barrier on
top of a layer of Ag is essentially independent of the substrate underneath.

The prefactor for strained
Ag on Ag\,(111) using transition state theory within the harmonic approximation
[cf. Eq. (\ref{harmappr})] has also been calculated~\cite{rat97_2}.  
In this calculation up to $99$ degrees of freedom were considered,
and it is interesting to note that within a factor of $2$ the result agreed
with that obtained by taking into account only the degrees of freedom of
the moving adatom.
It was found that the prefactor changes only very little
when the lattice constant (and thus the diffusion barrier, cf.
Fig.~\ref{Strain_barrier}) changes, so that there is no significant
compensation effect.  The results are summarized in Table \ref{prefactors}.
\begin{table}[tb]
\vspace{.5cm}
\caption{
  Prefactor $\Gamma_0$ (in THz) and energy barrier $E_{\rm d}$ (in meV) for
  strained Ag on Ag\,(111), and for Ag on one monolayer (ML) Ag on Pt\,(111). The
  prefactors are calculated within the harmonic approximation of transition
  state theory. All results shown were obtained by considering 15 degrees of freedom.}
\begin{tabular}{lcccc}
\hline
Substrate & Ag\,(111) & Ag\,(111) & Ag\,(111) & 1 ML Ag on Pt\,(111) \\
& $a=3.92$ \AA & $a=4.05$ \AA & $a=4.22$ \AA & $a=3.92$ \AA \\
\hline
$\Gamma_0$  & 1.3 & 0.8 & 0.3 & 7.1 \\
$E_{\rm d}$ & 60 & 81 & 105 & 63 \\
\hline
\end{tabular}
\label{prefactors}
\end{table}
All the values calculated are of the order of 1 THz, as 
expected, since this is a typical surface phonon frequency. Thus, 
the calculations do not confirm the claim by the authors of Ref. \cite{bru95}
that there is a very large compensation effect for this system.
An additional test has been done 
to investigate whether the prefactor (in contrast to the diffusion 
barrier) might be affected by the substrate underneath, but calculations 
for the prefactor for Ag on 1 monolayer Ag on Pt\,(111) also gave a 
prefactor of approximately 1 THz (cf. Table \ref{prefactors}).

The reason for the large discrepancy between the theoretical and the experimentally 
deduced value for the prefactor for the system under compressive 
strain is not completely understood yet. We believe that the 
calculations are correct, and that the interpretation of the experimental 
data may be flawed. Possible reasons for this have been given in the 
previous Section when we introduced Eq.~(\ref{scaling}). 
A low diffusion 
barrier implies that the PES has a very small corrugation so that 
it is likely that small clusters diffuse.
Also, long range effects (such as reconstructions or 
strain fields) may become important. 
The quantity $C_0$ in Eq.~(\ref{scaling}) depends on the capture 
numbers $\kappa_j$, but it 
is not clear how the $\kappa_j$ behave in a strained system.
We thus conclude that one has to be rather careful if one wants to 
use simple scaling laws to determine 
microscopic quantities from experimental data.

\subsection{Hopping and Exchange Diffusion on (100) Surfaces}

As it has been pointed out already in Section~\ref{sec:atomistic}, diffusion
may occur by hopping or atomic exchange on the (100) surface.
Fig.~\ref{more_exchange} shows that the transition state of the exchange
mechanism can be described by a dimer, and each atom of this dimer can form
three chemical bonds, so that a simple bond counting argument suggests that
the exchange mechanism is always the energetically preferred mechanism.  It
has been calculated by Feibelman~\cite{fei90} that the covalent nature of
aluminum which allows the formation of directional bonds at certain geometries
lowers the total energy of the system in the transition state for atomic
exchange.  Thus, the desire of the system to maximize the coordination of all
atoms involved is the reason why diffusion proceeds via atomic exchange.

This is certainly a plausible argument for Al\,(100), but in general chemical
bonding is a lot more complicated than counting nearest neighbor bonds and
this explanation can not account for atomic exchange as the energetically
preferred mechanism for Pt and Ir.  Moreover, it is puzzling that for the
transition metals Ir and Pt diffusion proceeds via exchange, but according to
calculations by Yu and Scheffler ~\cite{yu96} for Ag the preferred diffusion
mechanism is hopping. The authors of Ref.~\cite{yu96} find that the barrier
for hopping is $0.52$~eV, while that for exchange is $0.93$~eV when the
exchange-correlation energy is approximated within the LDA (these numbers
change to $0.45$~eV and $0.73$~eV within the GGA).

In a subsequent paper~\cite{yu97}, Yu and Scheffler provided the following
explanation for the different behavior of different transition metals. They
realized that for $5d$ metals (Ir and Pt, with preferred exchange mechanism)
the surface stress is significantly higher than it is for 3$d$ and 4$d$ metals (Ag,
where hopping is energetically favored)~\cite{fio93}.  This is caused by relativistic
effects that play a more important role for the heavier 5$d$ metals.  Higher
tensile surface stress has a tendency to pull the dimer in the transition
state for atomic exchange closer toward the surface and the layer underneath
to lower the total energy of the system. Thus, under tensile surface stress
the barrier for atomic exchange decreases, in contrast to the barrier for
hopping, that in analogy to the behavior on the (111) surface is expected to
increase under tensile surface stress.

\begin{figure}[tb]
\unitlength1cm
\begin{center}
   \begin{picture}(7,5.5)
      \includegraphics{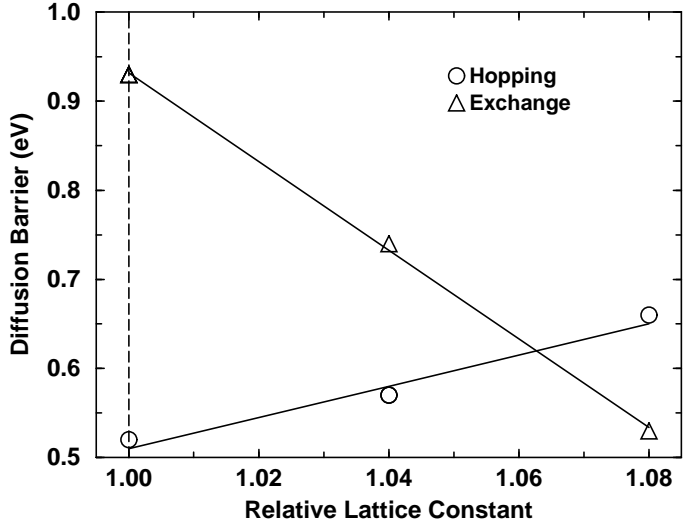}
   \end{picture}
\end{center}
\caption{
  The energy barrier for the hopping and the exchange mechanism for Ag on
  Ag\,(100) as a function of the relative lattice constant $a/a_0$.}
\label{more_strain_barrier}
\end{figure}

To test this claim, Yu and Scheffler calculated the energy barrier for
hopping and exchange as a function of the lattice constant for Ag on Ag\,(100),
and the results are shown in 
Fig.~\ref{more_strain_barrier}.
Indeed, the two curves cross at approximately $6.5\%$ lattice
mismatch, and it is predicted that under high tensile surface stress the
energetically preferred mechanism is exchange.  Yu and Scheffler also claimed
that for unreconstructed Au\,(100) the energetically preferred mechanism should
be exchange, with a barrier of $0.65$~eV (compared to $0.83$~eV for hopping).
This again is due to the fact that the dimer in the transition state of the
the exchange mechanism is attracted to the surface; it is approximately $37\%$
closer to the surface than the interlayer spacing in the bulk would suggest.
This is similar to Al\,(100), where the dimer is $55\%$ closer to the surface,
but in contrast to the (unstrained) Ag\,(100), where the dimer is only $25\%$
closer to the surface.  When Ag\,(111) is strained by $8\%$ the dimer is $57\%$
closer to the surface.

It seems that there is clear evidence that atomic exchange is stabilized by
the correlation of the local coordination of all atoms involved and the
strength of the bonds (as for Al), and by the tensile surface stress. Following
this argument, the opposite should happen under compressive surface stress:
When the surface is compressed, the dimer is ``squeezed'' away from the
surface, and as a result the energy barrier for exchange increases. This is
again in contrast to the behavior of the barrier for hopping, that is expected
to decrease. Recent results for Al on Al\,(100)~\cite{bos97}  show that indeed the 
energetically preferred diffusion mechanism changes from exchange to hopping under 
compressive strain.

Before closing this Section, we would like to point out that based on
knowledge for the barrier for surface diffusion it is not appropriate to
conclude that one mechanism is the preferred mechanism. The diffusion constant
also depends on the prefactor, and there are indications in the
literature~\cite{liu91} that the prefactor for diffusion by exchange might be
1 or 2 orders of magnitude larger than the prefactor for hopping. In fact, the
consequences of this are discussed in more detail in the example in the next
Section. From the results for diffusion of Ag on Ag\,(111) it can be expected
that there is not a significant strain dependence of the prefactor, but more
research should be directed toward a better knowledge of the different
prefactors for different mechanisms and different surfaces.

\section{AB INITIO KINETIC MONTE CARLO\\ SIMULATIONS}
\label{sec:kmc}

The individual microscopic processes can be used to get a deeper understanding
of epitaxial growth, provided that their interplay as a function of temperature
and deposition rate can be modeled properly.  The kinetics during growth is a
stochastic process that is well described by a master equation, and the Monte
Carlo technique yields a numerical solution of this
equation~\cite{MC_references1,MC_references2}. The master equation governs the
dynamic evolution of $P({\bf{\sigma}},t)$, the probability that the system is
in state ${\bf{\sigma}}$ at time $t$, and it has the following general form:
\begin{equation}
\frac{\partial P({\bf{\sigma}},t)}{\partial t} = - \sum_{{\bf{\sigma '}}}
W({\bf{\sigma}} \to {\bf{\sigma '}}) P({\bf{\sigma}},t) + \sum_{{\bf{\sigma}}}
W({\bf{\sigma '}} \to {\bf{\sigma}}) P({\bf{\sigma '}},t) \quad .
\label{master}
\end{equation}
Here $W({\bf{\sigma '}} \to {\bf{\sigma}})$ has the meaning of a transition
probability from state ${\bf{\sigma '}}$ to state ${\bf{\sigma}}$ per unit
time. On the right-hand side the first sum describes the rate of all processes
where the system jumps out of the considered state (and hence decreases its
probability), while the second sum describes the rate of all processes where
the system jumps into the considered state (and hence increases its
probability). The solution of the master equation is achieved computationally
by choosing randomly among several possible transitions and accepting or
refusing the particular events according to an appropriate probability.
In thermal equilibrium the detailed balance condition
\begin{equation}
W({\bf{\sigma}} \to {\bf{\sigma '}}) P_{\rm eq}({\bf{\sigma}}) =
W({\bf{\sigma '}} \to {\bf{\sigma}}) P_{\rm eq}({\bf{\sigma '}})
\label{detbal}
\end{equation}
ensures that the two sums on the right-hand side of Eq. (\ref{master}) cancel
and $P_{\rm eq}({\bf{\sigma}})$ is the steady-state distribution of the master
equation.

The obvious question concerns the choice of a pertinent transition
probability $W({\bf{\sigma}} \to {\bf{\sigma '}})$. The detailed balance
criterion, however, only determines the ratio $W({\bf{\sigma}} \to {\bf{\sigma
    '}})/W({\bf{\sigma '}} \to {\bf{\sigma}})$. When static properties of the
system are sought, the Metropolis algorithm~\cite{met53} is appropriate.
According to this scheme the probability that a new configuration is accepted
is proportional to $\exp(-\Delta E/k_{\rm B}{\rm T})$, where $\Delta E$ is the
difference between the total energies of the system in the new and old
configuration.  Such an algorithm searches for the configuration corresponding
to the minimum of the total energy. The sequence of generated configurations
does not correspond to the real time evolution of the system, and there is no
problem of conversion from simulated time $t_s$ to real time $t_r$. However,
for the dynamic behavior (such as growth) of a system the correlation between
$t_s$ and $t_r$ is important, and the configuration sequence has to mimic the
real one. The appropriate tool for this is KMC.  Here the dynamics of the
system is obtained from the local microscopic transition rates in terms of
Poisson processes.

In contrast to MD, in KMC the place and time of events are
no longer deterministically obtained, but are chosen by statistical
considerations. The task of the KMC is to build up an artificial chronological
sequence of distinct events separated by certain {\it interevent} times. This
chronology is based on the hierarchy dictated by the transition rates and may
not correspond to reality.  For example, during growth events may occur
simultaneously, but KMC introduces an artificial time interval between them.
At the other extreme, events may be separated by a very large interval that is
not reflected in the KMC {\it interevent} time.  Thus, KMC does not model the
``deterministic'' microscopic dynamics yielding the exact times of various
processes (as MD does). Therefore, in KMC it is possible to avoid the explicit
calculation of all unsuccessful attempts, but the chain of events and
corresponding interevent times must be constructed from probability
distributions weighting appropriately all possible outcomes.

To illustrate the strategy of a KMC simulation (for more details see
Ref.~\cite{woodru97}) let us consider a system of $M$
particles that are capable of undergoing o total of $m$ transition events
(diffusions, deposition, etc.). The $m$ transition events are described by a
set of rates ${\bf{\Gamma}} \equiv \{\Gamma^{(1)},
\Gamma^{(2)},.....,\Gamma^{(m)}\}$ where $\Gamma^{(j)}$ has the form given by
Eq. (\ref{retdef1}). The $M$ particles can be partitioned in classes among the
various possible transition events as ${\bf M} \equiv \{n^{(1)},
n^{(2)},.....,n^{(m)}\}$ where $n^{(j)}$ is the number of particles capable of
undergoing a transition with a rate $\Gamma^{(j)}$. It should be pointed out
that $\sum_j n^{(j)} \geq M$ since a particle may belong to several classes of
${\bf M}$ at the same time. Thus, a configuration of the system at a
particular time can be labeled by the distribution of ${\bf M}$ over
${\bf{\Gamma}}$. A transition is determined by picking randomly among various
possible events available at each time step, and the configuration is updated.
The choice of the event is dictated by the dynamical hierarchy described by
the set of relative transition probabilities {\bf W} $\equiv \{W^{(1)},
W^{(2)},.....,W^{(m)}\}$, each of them defined as:
\begin{equation}
  W^{(j)} = \frac{n^{(j)}\,\Gamma^{(j)}}{R} \quad ,
\label{relatprob}
\end{equation} 
where $R = \sum_j n^{(j)}\,\Gamma^{(j)}$ is the total rate.
These probabilities are constructed such that detailed balance is achieved at
thermal equilibrium. After each Monte Carlo step time should be updated with
an increment $\Delta t = -ln(\rho)/R$ where $\rho$ is a random number in the
range $(0,1]$. In this way, a direct and unambiguous relationship between KMC
time and real time is established, since the KMC algorithm effectively
simulates stochastic processes described by a Poisson distribution.

KMC simulations have been used to study crystal growth of semiconductors
(e.g.~\cite{semimc1,semimc2,semimc3}) and metals
(e.g.~\cite{work1,work2,work3,work4}).  However, most of these studies have
been based on restrictive approximations.  For example, the input parameters
have been treated as effective parameters determined rather indirectly by
fitting to experimental quantities, like intensity oscillations in helium atom
scattering (HAS) measurements or reflection high energy electron diffraction
(RHEED), or they were obtained from STM studies of island densities. Thus, the
connection between these parameters and the microscopic nature of the
processes may be somewhat uncertain.  Often the correct surface structure was
neglected and the simulation was done on a simple cubic lattice while the
system of interest had an fcc or bcc structure.  Despite these approximations
such studies have provided qualitative and in some cases also quantitative
insight into growth phenomena. 

It is desirable to carry out KMC simulations
with the proper geometry and microscopically well founded parameters. This has
been done for example in Refs.~\cite{liu93,jac95,sh96} where semi-empirical
calculations have been employed to evaluate the PES. The most
accurate, but also most elaborate approach to obtain the PES employs DFT as
described in Sections~\ref{sec:concepts} and \ref{sec:atomistic}. It is interesting
to note that the result of a KMC study will be the same as that of a MD
simulation, provided that the underlying PES is the same. More details on the
KMC method are contained in Ref.~\cite{woodru97} and references therein.

An example of the possible combination of DFT results and KMC simulations is
the study of island shape in functions of substrate temperature for Al on
Al\,(111). This surface, as each (111) surface of an fcc crystal, is
characterized by the presence of two types of close-packed steps, labeled as
$\{100\}$ and $\{111\}$ facets, referring to the plane passing through the
atoms of the step and the atom of the substrate (often these steps are labeled
A and B). Using the DFT parameters for Al/Al\,(111) by Stumpf and
Scheffler~\cite{stu94,stu96} one can carry out KMC simulations under typical growth
conditions. 
\begin{figure}[tb]
  \leavevmode
%  \special{psfile=pagis7 hoffset=-25 voffset=-705
  \includegraphics{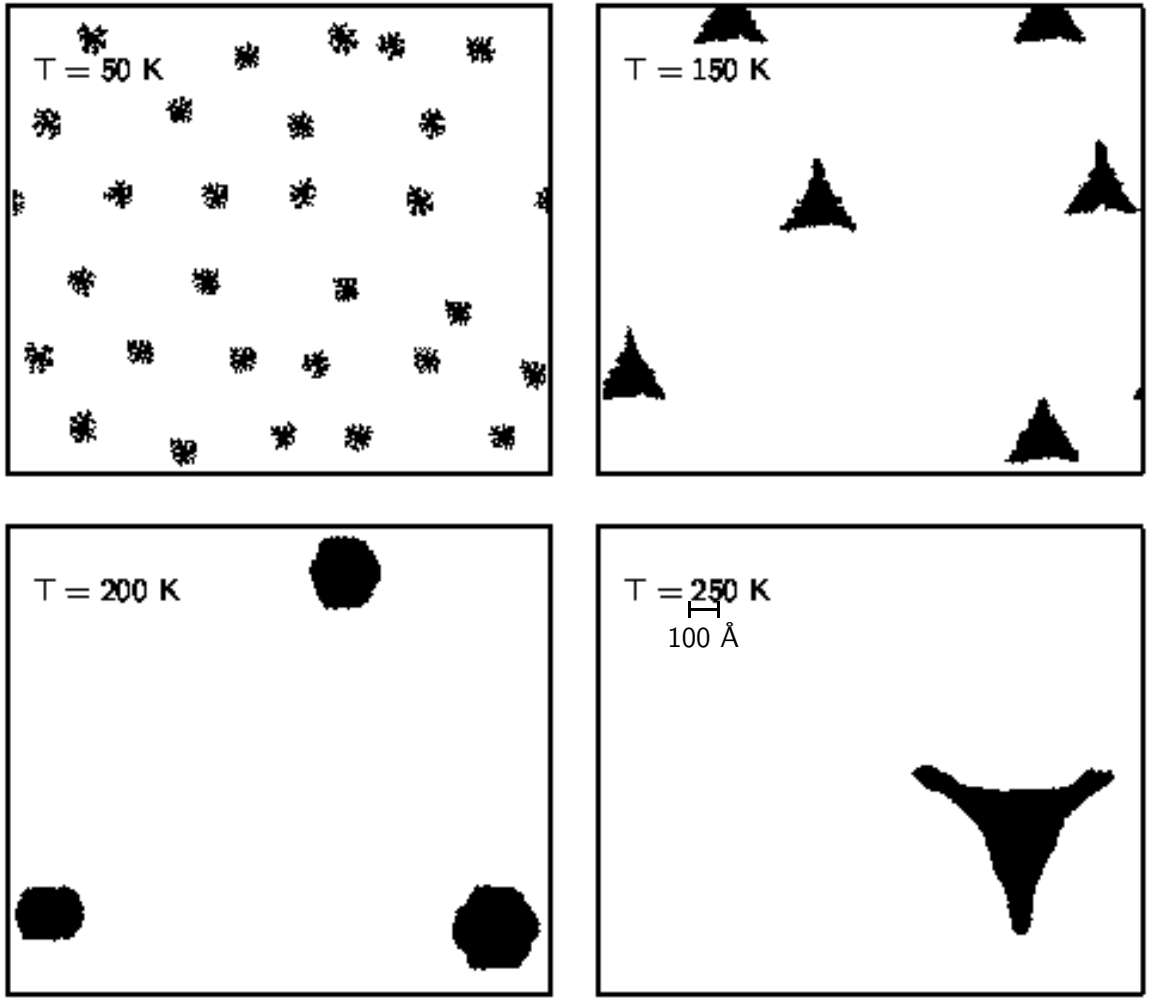}
\unitlength1cm
\begin{picture}(0,0)
\thicklines
\put(9,-8){\line(1,0){0.38}}
\put(9,-8.1){\line(0,1){0.2}}
\put(9.38,-8.1){\line(0,1){0.2}}
\put(8.7,-8.5){\small{\textsf{100 \AA}}}
\end{picture}
\vspace*{13.0cm}
\caption{
  A surface area of (1718 $\times$ 1488)~\AA$^2$~ (half of the simulation
  area) at four different substrate temperatures.  The deposition rate was
  0.08 ML/s and the coverage in each picture is $\Theta$ = 0.04 ML.}
\label{fig.2}
\end{figure} 
The results of the {\it ab initio} KMC simulations where all
relevant processes, including the different mechanisms for a particular event
(i.e. hopping or exchange) are included, are shown in Fig.~\ref{fig.2}. They
are obtained after 0.5 s with a deposition rate of 0.8 ML/s. When the growth
temperature is 50 K the shape of the islands is highly irregular and indeed
fractal. Adatoms which reach a step cannot leave it anymore and they cannot
even diffuse along the step.  Thus, at this temperature ramification takes
place into random directions, and island formation can be understood in terms
of the hit and stick model.  At $T$ = 150 K the island shapes are triangular
with their sides being \{100\}-faceted steps, but increasing the temperature
to $T$ = 200 K a transition from triangular to hexagonal shape occurs. For $T$
= 250 K the islands become triangular again, but they are rotated by $\pi$ and
are mainly bounded by \{111\}-faceted steps.

The transition can be understood in terms of different diffusivities along the
step edges of the islands: The lower the migration probability along a given
step edge, the higher is the step roughness and the faster is the speed of
advancement of this step edge. As a consequence, this step edge shortens and
eventually it may even disappears.  Since diffusion along the densely packed
steps on the (111) surface (the \{100\} and \{111\} facets) is faster than
along steps with any other orientation, this criterion explains the presence of
islands which are mainly bounded by \{100\}- or \{111\}-faceted steps. The
same argument can be extended to the diffusion along the two close-packed
steps. By considering the energy barriers (0.32 eV for diffusion along the
\{100\} facet vs. 0.42 eV along the \{111\}) we would expect only islands with
\{100\} sides, until the temperature regime for the thermal equilibrium is
reached. This is the case for $T$ = 150 K where the energy barrier dominates
the diffusion rate.

DFT calculations~\cite{stu94} have shown that an adatom migrates along the
\{100\} facet by hopping, while along the \{111\} the diffusion occurs via the
exchange mechanism.  Since the attempt frequency for exchange is expected to
be higher than the one for hopping (in this simulation it is assumed to be two
orders of magnitude larger), the higher $\Gamma_0^{\{111\}}$ leads to faster
diffusion along the \{111\} facet than along the \{100\} one at $T$ = 250 K.
The latter steps disappear and only triangles with \{111\}-faceted sides are
present. The roughly hexagonally shaped islands at $T$ = 200 K are a
consequence of the equal advancement speed for the two steps at that
temperature.  A similar transition in the island shapes has been observed for
Pt\,(111)~\cite{mic93}. An important point regards the time required by the
system to assemble these structures. For example, with $F = 0.8$ ML/s the
island at $T$ = 250 K assumes a triangular shape for $t \geq 0.1$ s, a time
not reachable by MD simulations.

In conclusion, during growth the crystal surfaces may show many structural
details such as islands, clusters, steps, all of which are originated on an
atomic length scale (microscopically well-founded description is needed) but
may lead to cooperative phenomena of mesoscopic dimensions with time scale of
the order of seconds (stochastic methods are necessary).

\section{ACKNOWLEDGMENTS}
The authors are indebted to M. Fuchs, A. Kley, A.P. Seitsonen, and B.D. Yu for
many valuable discussions.


\begin{thebibliography}{999}
\bibitem{units} 
All equations are noted in Hartree atomic units, i.e.,
the unit of length is 1 bohr = 0.5292 \AA, the unit of energy
is 1 hartree = 27.2116 eV, and $\hbar = m = e = 1$.
\end{thebibliography}
\end{document}